\documentclass{ws-procs975x65}

\begin{document}



\title{ POSSIBLE NEUTRINO-ANTINEUTRINO OSCILLATION UNDER GRAVITY AND ITS CONSEQUENCES  
}

\author{BANIBRATA MUKHOPADHYAY
}

\address{Department of Physics,
Indian Institute of Science, 
Bangalore-560012, India 
\email{bm@physics.iisc.ernet.in}}

%


\def\lsim{\lower.5ex\hbox{$\; \buildrel < \over \sim \;$}}
\def\gsim{\lower.5ex\hbox{$\; \buildrel > \over \sim \;$}}

\def\ch{\lower-0.55ex\hbox{--}\kern-0.55em{\lower0.15ex\hbox{$h$}}}
\def\lh{\lower-0.55ex\hbox{--}\kern-0.55em{\lower0.15ex\hbox{$\lambda$}}}

\begin{abstract}
We show that under gravity the effective masses for neutrino and antineutrino are
different which opens a possible window of neutrino-antineutrino oscillation even
if the rest masses of the corresponding eigenstates are same. This is due to CPT
violation and possible to demonstrate if the neutrino mass eigenstates are expressed
as a combination of neutrino and antineutrino eigenstates, as of the neutral kaon
system, with the plausible breaking of lepton number conservation. In early universe,
in presence of various lepton number violating processes, this oscillation might lead
to neutrino-antineutrino asymmetry which resulted baryogenesis from the B-L symmetry
by electro-weak sphaleron processes. On the other hand, for Majorana neutrinos,
this oscillation is expected to affect the inner edge of neutrino dominated accretion
disks around a compact object by influencing the neutrino sphere which controls the
accretion dynamics, and then the related type-II supernova evolution and the r-process
nucleosynthesis.
\end{abstract}

\bodymatter

\section{Introduction}\label{intro}

\enlargethispage*{6pt}

The neutrino oscillation, in the flat space, is due to difference in rest masses
between two mass eigenstates. However, in late eighties, it was
first pointed out \cite{gas} that presence of gravitational field
affects different neutrino flavors differently which violates equivalence
principle and
thus governs oscillation, even if neutrinos are massless or of degenerate mass.
The neutrino oscillation 
with LSND data \cite{mansar} indeed can be explained by
degenerate or massless neutrinos with flavor non-diagonal gravitational coupling.
It was further argued \cite{ab} that the flavor oscillation is possible 
in weak gravitational field with the
probability phase proportional to the gravitomagnetic field.
The oscillation was also shown to be feasible
when the maximum
velocities of different neutrino differ each other, even if they
are massless \cite{cg}.

All the above results are for flavor oscillation
or/and without considering rigorous general relativistic effects. 
However, properties of neutrino in curved spacetime have already
been discussed \cite{schw,pal,mukh} in literature. 
Here we address the {\it neutrino$-$antineutrino oscillation},
which violates lepton number conservation,
focusing on the nature of space-time curvature and its special effect.

While the neutrino$-$antineutrino oscillation
under gravity is an interesting issue on its own right, the present
result is able to address two long-standing
mysteries in astrophysics and cosmology: (1) Source of abnormally large
neutron abundance to support the r-process nucleosynthesis
in astrophysical site. (2) Possible origin of baryogenesis.

\section{Oscillation probability }\label{prob}

Let us recall the fermion Lagrangian density in curved spacetime \cite{schw,mukh} 
\begin{eqnarray}
{\cal L}=\sqrt{-g}\,\overline{\psi}\left[(i\gamma^a\partial_a-m)+\gamma^a\gamma^5
B_a\right]\psi ={\cal L}_f+{\cal L}_I, ~~~
\label{lagf}
\end{eqnarray}
where
\begin{eqnarray}
B^d=\epsilon^{abcd} e_{b\lambda}\left(\partial_a e^\lambda_c+\Gamma^\lambda_{\alpha\mu}
e^\alpha_c e^\mu_a\right),
\,\,\,\, e^\alpha_a\, e^\beta_b\,\eta^{ab}=g^{\alpha\beta},
\label{bd}
\end{eqnarray}
the choice of unit is $c=\ch=k_B=1$. ${\cal L}_I$ may be a CPT violating interaction and thus
the corresponding dispersion energy \cite{mukh} for neutrino and 
antineutrino in standard model 
\begin{eqnarray}
E_{\nu} =  \sqrt{({\vec p} - {\vec B})^2 + m^2} + B_0, ~~~~
E_{\overline{\nu}} = \sqrt{({\vec p} + {\vec B})^2 + m^2} -
B_0. \label{edis}
\end{eqnarray}
Eq. (\ref{edis}) tells us that under gravity neutrino energy is split up from antineutrino energy.
The CPT status of ${\cal L}_I$ has been discussed in detail in our previous works \cite{mukh}.

Now motivated by the neutral kaon system, we consider two distinct orthonormal
eigenstates $|E_\nu>$ and $|E_{\overline \nu}>$ for a neutrino and an antineutrino type respectively.
Further we introduce a set of neutrino mass
eigenstates at $t=0$ as \cite{baren}
\begin{eqnarray}
|m_1>=cos\theta\, |E_\nu>+sin\theta\, |E_{\overline \nu}>,\hskip0.5cm
|m_2>=-sin\theta\, |E_\nu>+cos\theta\, |E_{\overline \nu}>.
\label{fl2}
\end{eqnarray}
Therefore,
in presence of gravity,
the oscillation probability for $|m_1(t)>$ at $t=0$
to $|m_2(t)>$ at a later time $t=t_f$ can be found as
\begin{eqnarray}
P_{12}
=sin^22\theta\,sin^2\delta,\,\,\,\,\,\,
\delta=\frac{(E_\nu-E_{\overline \nu})t_f}{2}=
\left[(B_0-|\vec{B}|)+\frac{\Delta m^2}{2|{\vec p}|}\right]\,t_f,
\label{pab}
\end{eqnarray}
where we consider ultra-relativistic neutrinos.
Normally, the rest mass difference of particle and antiparticle is zero and thus 
possible $\delta\neq 0$ is mostly due to $B_a\neq 0$ i.e. due to gravitational coupling.
Therefore, the neutrino-antineutrino oscillation may be possible in presence of gravity
provided there is a lepton number violating process. If neutrinos exhibit Majorana mass,
then lepton number violation is automatically taken care.
Hence, the CPT violating nature of background curvature coupling generates effective mass
difference, while lepton number violating process leads to oscillation between neutrino
and antineutrino.

The oscillation probability is maximum at $\theta=\pi/4$ and is zero at $\theta=0,\pi/2$.
From eqn. (\ref{pab}), the oscillation length, $L_{osc}$,
by appropriately setting dimensions, is obtained as
\begin{eqnarray}
L_{osc}= c\,t_f=\frac{\pi\,\ch\,c}{\tilde{B}}\sim\frac{6.3\times 10^{-19}GeV}{\tilde{B}}{\rm km},
\label{old1}
\end{eqnarray}
where $\tilde{B}=B_0-|\vec{B}|$ is expressed in GeV unit and
the neutrino is considered to be moving in the speed of light.


\section{Consequence and Discussion}\label{dicu}

One of the situations where the gravity induced neutrino-antineutrino oscillation 
may occur is the GUT era of anisotropic phase
of early universe when $\tilde{B}\sim 10^5$ GeV \cite{mukh}.
From eqn. (\ref{old1}), this leads to
$L_{osc}\sim 10^{-24}$km which is $10^{14}$ orders of magnitude larger than the
Planck length. This has an important implication as the size
of universe
at the GUT era is within $\sim 10^{26}$ times of the Planck. Therefore, the oscillation
may lead to leptogenesis and then to baryogenesis by electro-weak sphaleron processes
due to $B-L$ conservation, what we see today.

Another plausible region for an oscillation of this kind to occur
is the inner accretion disk of the neutrino dominated accretion flow (NDAF) \cite{ndaf} 
around a rotating compact object which can be extended 
upto several thousand Schwarzschild radius.
From eqns. (\ref{bd}) and (\ref{old1}) we can obtain 
\begin{eqnarray}
B^0 =-\frac{4a\sqrt{M}z}{\bar{\rho}^2\sqrt{2r^3}},\,\,\,\,\,
L_{osc}\sim \frac{1.8\,x^{7/2}\,M_s}{a\,H}{\rm km}=\frac{1.2\,x^{7/2}}{a\,H}M,
\label{b0z}
\end{eqnarray}
for the Kerr geometry,
where $\bar{\rho}^2=2r^2+a^2-x^2-y^2-z^2$. The detailed calculation and discussion are
presented elsewhere \cite{mukh2}.
Here we choose the mass of the compact object $M=M_s\,M_\odot$, radius and height of the disk orbit
where oscillation takes place respectively $r\sim\bar{\rho}= x\,M$ and $z=H\,M$, and
we assure $\tilde{B}\sim B_0$.  Any oscillation at the
inner edge of NDAF is expected to be influenced by gravity what affects 
the accretion dynamics and outflow.
From eqn. (\ref{b0z}), $L_{osc}$ varies from a few factors to several hundreds of
Schwarzschild radii at $x\le 10$ for a fast spinning compact object. 

Supernova is thought to be the astrophysical site of the r-process nucleosynthesis.
During supernova, neutron capture processes for radioactive elements take place
in presence of abnormally large neutron flux. However, how does the large neutron flux
arise is still an open question. There are two related reactions:
\begin{equation}
\,\,n+\nu_e\rightarrow p+e^-,\,\,\,\, \,\,\,\,\,p+\bar{\nu}_e\rightarrow n+e^+.
\label{nuc}
\end{equation}
If $\bar{\nu}_e$ is over abundant than $\nu_e$, then, from eqn. (\ref{nuc}), neutron production
is expected to be more than proton production into the system. Therefore, the possible
conversion of $\nu_e$ to $\bar{\nu}_e$ due to the gravity induced oscillation
explains the overabundance of neutron.






\end{document}